\documentclass[aps,prd,groupedaddress,nofootinbib,letterpaper]{revtex4}

\usepackage{amsmath}
\usepackage{amsfonts}
\usepackage{amssymb}
\usepackage{amsthm}

\usepackage{braket,mathtools}
\usepackage{mathrsfs}
\usepackage{stmaryrd}

\usepackage{hyperref}
\usepackage{color}

\newcommand{\calo}{{\cal O}}

\newcommand{\beq}{\begin{equation}}
\newcommand{\eeq}{\end{equation}}
\newcommand{\hf}{\frac{1}{2}}
\newcommand{\OO}{\mathcal{O}}

\begin{document}
\begin{titlepage}

\title{Observables, gravitational dressing, and obstructions to locality and subsystems}

\author{William Donnelly}
\email{donnelly@physics.ucsb.edu}
\affiliation{Department of Physics, University of California, Santa Barbara, CA 93106}

\author{Steven B. Giddings}
\email{giddings@physics.ucsb.edu}
\affiliation{Department of Physics, University of California, Santa Barbara, CA 93106}

\begin{abstract}

Quantum field theory  -- our basic framework for describing all nongravitational physics -- conflicts with general relativity: the latter precludes the standard definition of the former's essential principle of locality, in terms of commuting local observables.
We examine this conflict more carefully, by investigating implications of gauge (diffeomorphism) invariance for observables in gravity.
We prove a dressing theorem, showing that any operator with nonzero Poincar\'e charges, and in particular any compactly supported operator, in flat-spacetime quantum field theory must be gravitationally dressed once coupled to gravity, {\it i.e.}\ it must depend on the metric at arbitrarily long distances, and we put lower bounds on this nonlocal dependence.  
This departure from standard locality occurs in the most severe way possible: in perturbation theory about flat spacetime, at leading order in Newton's constant.
The physical observables in a gravitational theory therefore do not organize themselves into local commuting subalgebras: the principle of locality must apparently be reformulated or abandoned, and in fact we lack a clear definition of the coarser and more basic notion of a quantum subsystem of the Universe.
We discuss relational approaches to locality based on diffeomorphism-invariant nonlocal operators, and reinforce arguments that any such locality is {\it state-dependent} and {\it approximate}.
We also find limitations to the utility of bilocal diffeomorphism-invariant operators that are considered in cosmological contexts.
An appendix provides a concise review of the canonical covariant formalism for gravity, instrumental in the discussion of Poincar\'e charges and their associated long-range fields.
\end{abstract}

\maketitle

\end{titlepage}

\section{Introduction}

A complete theory of quantum physics must include gravity, yet certain features of gravity, already apparent even in the weak-field regime, appear to imply a significant departure from the traditional approach to physics via local quantum field theory (LQFT).  Specifically, the foundational principles of quantum field theory are the principles of quantum mechanics, of special relativity, and of locality.  While na\"ively gravity merely involves generalization of the principles of relativity, a more complete understanding reveals a more profound challenge to the foundations of local quantum field theory: once the invariance of the theory includes general coordinate transformations (or some precursor symmetry in a more fundamental theory), the very definition or formulation of locality is no longer apparent.  

Indeed, a precise formulation of locality of local quantum field theory is most clearly given in the algebraic approach\cite{Haag} to the subject, where locality is described in terms of commutativity of subalgebras of observables associated with spacelike separated regions of spacetime.
This results in a ``net" structure of subalgebras of the algebra of observables, which mirrors the topological structure of open sets 
of the spacetime manifold.\footnote{In fact, the net structure of the algebra encodes more than just the topology: it encodes the causal structure, which in turn determines the metric up to a conformal factor. 
While fluctuations in the causal structure are an obstruction to defining commuting subalgebras in gravity, this is an essentially nonperturbative effect.
The effects we will discuss are more severe in that they arise in perturbation theory. }
However, if one considers gravity, the requirement that observables be gauge invariant then serves as a major obstacle to an analogous definition of locality\cite{SGalg,DoGi}.  

In fact, an even more primitive, and typically essential, structure for physics is a decomposition of a physical system into {\it subsystems}.  In finite quantum systems, this is usually described through a tensor factorization of the Hilbert space.  In quantum field theory, 
the algebraic net structure provides an alternative to tensor factorization as a way of defining  subsystems that are localized.  
Tensor  factorization is generally problematic for two essentially independent reasons.
The first problem is that the algebras associated with regions of space in field theory are  type III von Neumann algebras, and do not correspond to the full algebra of bounded operators on any Hilbert space.  This means that the Hilbert space cannot be factored into factors corresponding to adjacent regions.  
The issue is essentially ultraviolet in nature (infinite entanglement of short-distance degrees of freedom), so may plausibly be resolved -- though in an unknown way -- in quantum gravity, where we do not trust LQFT to arbitrarily high energies.  In the absence of such a resolution, this issue can be avoided by a definition of subsystems based on the commuting subalgebras.   But an apparently  more severe issue is the presence of constraints associated with gauge invariance.  
Specifically, as we will describe, these constraints require long-range gravitational fields, which obstruct commutativity of operators, and thus interfere with an algebraic definition of subsystems via commuting subalgebras.  Since a tensor factorization would imply commutativity of operators defined within separate factors, this also prevents tensor factorization, though the algebraic structure is the more general structure.
This is essentially an infrared effect and we do not expect it to be solved even given a gauge-invariant UV regulator.   

Problems with locality and observables in gravity have been known for some time.  For example, \cite{Torre:1993fq} argued that, in a closed universe, there are no observables that are local in time, by considering general integrals over a Cauchy surface of any 3-form constructed from the 3-metric, its canonical momentum, and finitely many derivatives thereof.  One of the goals of this paper is to sharpen and extend such statements.  As an example, in a LQFT of a scalar field, the basic local observable is the field operator, which creates or annihilates a particle.  When we couple to gravity we still expect that there are operators that create or annihilate particles.  But, for these obervables to be gauge invariant, they must also be ``gravitationally dressed," so that they create or annihilate the gravitational field configuration of the particle; put differently, a particle is inseparable from its gravitational field.  This dressing makes these observables nonlocal, as was explicitly described in \cite{SGalg,DoGi}.

We will explore the requirement of gravitational dressing in more generality in this paper.  In particular, one might ask whether such dressing can be screened, as is possible in gauge theories of internal symmetries, such as quantum electrodynamics.  In these theories, this then permits the definition of local commuting subalgebras, which provide a way of defining locality for such theories.  After considering  examples, we prove a general result: an operator with nonzero Poincar\'e charges (momentum, angular momentum) must  act nontrivially on the asymptotic gravitational field, and so its gauge-invariant, gravitationally dressed version must be nonlocal.  This severely restricts the possible definition of local subalgebras and a corresponding algebraic definition of subsystems.

One can then ask what operators one can find with vanishing Poincar\'e charges.  In the classical limit, positivity of the energy indicates that positive energy of matter cannot, for example, be screened by some negative gravitational energy.  The simplest examples of Poincar\'e-singlet operators are integrals over all of $D$-dimensional spacetime of $D$-form local operators, or, if one allows metric dependence, of scalar operators times the volume form.  Such ``single-integral" observables (in the nomenclature of \cite{GMH}) have a long history in the literature, see \cite{Dewi1,Maro}, and for earlier related work, \cite{GeDe1,GeDe2,Komar, BeKo}.  These integrated operators are completely nonlocal, although in certain states they approximately reduce to local operators.  So, as we discuss further, if one attempts to define locality via such operators, this locality is {\it state-dependent} and {\it approximate}.  So far, we don't know a better definition of locality in gravitational theories.  

In line with the comments above, the situation is worse, in that the trouble extends to that of even defining a coarser structure than that of locality -- we apparently lack a precise definition of quantum subsystems of our quantum system, the ``Universe," that reduces to the familiar notion of subsystems, based on local commuting subalgebras of LQFT, in the weak-gravity limit.  
In short, the basic properties of gravity, evident in the weak-field limit, appear to be at odds with the fundamental structure of local quantum field theory, our otherwise very successful framework for accurately describing the rest of physics.  
This is a conundrum which reaches beyond more  technical issues, such as those of specific UV or IR divergences, and plausibly has bearing on other deep puzzles such as the unitarity crisis for black hole evolution.  
Locality, and even the definition of subsystems -- basic elements of local quantum field theory and quantum mechanics, respectively -- are, unless defined in a fundamentally new way, 
both state dependent and approximate.

In our discussion we also examine a different type of observable that has been considered in gravity, where local operators are connected by a geodesic of fixed length. Such observables are candidate diffeomorphism-invariant observables that have been for example investigated in cosmological settings.  We find that while such operators can also be constructed in Minkowski backgrounds, they are, at least to leading order in the gravitational coupling, rather trivial:  at least in a free theory, they essentially reduce to just the number operator.

In outline, the next section (\ref{section:gauge-vs-locality}) gives a more extensive discussion of the question of screening, in comparison with gauge theory, and gives via examples an introduction to the necessity of vanishing Poincar\'e charges in order to avoid asymptotic gravitational dressing.  
This discussion naturally incorporates that of diffeomorphism-invariant bilinear operators and their leading-order triviality.  Section \ref{section:dress} then states and proves our result that nonvanishing Poincar\'e charges imply such nonlocal dressing.  
It also discusses the role of the positive energy theorem in precluding screening, and the consequences of such no-screening statements for the structure of the algebra of observables and for  locality.  
Section \ref{section:other} gives a discussion of the Poincar\'e-invariant single-integral observables, and their possible role in defining state-dependent, approximate locality, as well as briefly discussing an approach\cite{Donnelly:2016auv} where the theory is supplemented by extra degrees of freedom which are proposed to define a ``localization."  
Following the discussion and conclusions, we provide two appendices, \ref{appendix:conventions}  with conventions, and \ref{appendix:canonical} that gives a concise, self-contained account of the covariant canonical formalism for gravity.  
This formalism is used to understand the structure of the Poincar\'e charges that enter into the no-screening arguments of section \ref{section:dress}.

\section{Gauge invariance vs. locality and subsystems}
\label{section:gauge-vs-locality}

An important question for quantum gravity is how it reproduces the structure of local quantum field theory (LQFT), in the weak-gravity approximation.  In LQFT, locality is a fundamental principle.  Indeed, in an axiomatic approach to the subject, such as that of algebraic quantum field theory (or, ``local quantum physics;" see, {\it e.g.} \cite{Haag}), one of the basic postulates is the existence of subalgebras of operators that are associated with different regions of spacetime; two subalgebras that are associated with spacelike-separated regions commute.  More generally, a division of a system like the Universe into subsystems is a key ingredient for physics, and in LQFT these commuting subalgebras furnish such decompositions.  

While there has been a lot of discussion in gravitational theories of physical phenomena and quantities that assume the existence of such a subsystem decomposition (and, moreover, a factorization of the Hilbert space), such as entanglement, entropy, and quantum information transfer, so far we don't know how to define this more basic (and prerequisite) notion of subsystems in gravitational theories \cite{SGalg}.  Given the key role of commuting subalgebras of observables in defining such a structure for LQFT, it is natural to start by investigating the possibility of an analogous definition in gravity.  However, the requirement of gauge invariance produces a significant obstacle, as has been explored in \cite{SGalg,DoGi}.  Interestingly, the analogous obstacle does {\it not} arise in gauge theories of internal symmetries, such as in quantum electrodynamics (QED) or quantum chromodynamics (QCD).

Let us first consider this case of Yang-Mills theories.  Of course, gauge-invariant local operators such as ${\rm Tr}\, F^2(x)$ can be immediately written down.  
The question becomes more complicated when we consider operators that create charged particles; for simplicity let us consider a single charged scalar $\phi(x)$ coupled to QED.  The operator $\phi(x)$ itself is not gauge-invariant, but it is possible to construct gauge-invariant 
dressed operators that create charged excitations along with their electromagnetic fields \cite{Dirac1955}.
The dressing  for a single particle necessarily extends to infinity, and hence such an operator cannot be local.
This happens because the operator $\phi$ carries a nonzero charge;
since, as seen from Gauss' law, the charge is measurable at infinity, the operator must have a nontrivial effect on the field that extends to spatial infinity.  A charged particle cannot be separated from its electromagnetic field.

While this field is inevitable, it does not need to extend to infinity:  it can be screened by introducing a second operator creating an oppositely-charged particle.  Two such operators can be combined with an electromagnetic dressing that does not extend to infinity, such as
\begin{equation} \label{QEDline}
\phi(x) e^{iq \int_\Gamma dx^\mu A_\mu} \phi^*(x')\ ,
\end{equation}
where $\Gamma$ is a curve connecting $x$ and $x'$.  
If $U$ is a connected open set containing $x$ and $x'$, a $\Gamma$ can be found in $U$, and this {\it localized} operator will commute with analogous such operators associated with spacelike-separated open sets.  So, local subalgebras of such operators can be constructed for QED, and more generally for nonabelian theories.

In exploring the possibility of localized operators in gravity, a first question is whether there is an analogous construction there.  As in the QED case, a scalar field operator $\phi(x)$, coupled to gravity, is not gauge invariant.  Gauge-invariant operators can be found by solving the constraints, which tell us that such an operator must be gravitationally dressed.  The nature of this dressing may be explored to leading order in an expansion in the gravitational coupling constant, 
$\kappa = \sqrt{32 \pi G}$. We will work throughout in $D > 3$ spacetime dimensions, unless otherwise specified.\footnote{As is well known, the case $D=3$ has certain peculiar features\cite{GAK}.}
Gauge-invariant, dressed operators may be found\cite{DoGi} which are of the form
\begin{equation}\label{dressphi}
\Phi(x) = \phi(x) + V^\mu(x) \partial_\mu \phi(x) + O(\kappa^2), 
\end{equation}
where the vector field $V^\mu(x)$ is a choice of gravitational dressing, which is of order $\kappa^1$ and is defined as an integral of the linearized metric perturbation $h_{\mu \nu}$, defined by $g_{\mu \nu} = \eta_{\mu \nu} + \kappa h_{\mu \nu}$.
The simplest case of such a dressing is a ``gravitational Wilson line'' extending to infinity\cite{DoGi}, which is in many ways analogous to its electromagnetic counterpart, the Faraday line of \cite{Dirac1955}.
For this gravitational Wilson line observable the dressing is given by
\begin{align}
V_{W_z,z}(x) &= \frac{\kappa}{2} \int_0^\infty ds\, h_{zz}(x+ s \hat z), \nonumber \\
V_{W_z,\check \mu}(x) &= \kappa \int_0^\infty ds\left[h_{z \check \mu}(x+s \hat z) +\hf\partial_{\check \mu} \int_{s}^\infty ds'\, h_{zz}(x+s' \hat z)\right] \label{VWz},
\end{align}
where $\check \mu$ indicates the indices other than $z$.
Then, the operator  \eqref{dressphi}  is invariant under the linearized diffeomorphism symmetry.
Acting on the vacuum, the operator creates a scalar field excitation together with a linearized gravitational field concentrated on the line $\{x + s \hat z : s \geq 0 \}$.  This operator may also be averaged over angles to give the more standard ``Coulomb" field\cite{DoGi}.  In either case, the operator $\Phi(x)$ is nonlocal, with support extending to infinity.

This construction seems analogous to that for QED, and one might ask whether two such operators could be combined to create a nonlocal, but localized, operator, where the asymptotic gravitational dressing cancels.  This would provide an example of a gravitational analog of screening.
The simplest way to construct a diffeomorphism-invariant dressing of an operator such as $\phi(x) \phi(x')$ is simply to take a product of diffeomorphism-invariant operators:
\begin{equation}\label{bidef}
\mathcal{A} = \Phi_{W_z}(x) \, \Phi_{W_z}(x') \ .
\end{equation}
In the analogous QED case, for equal but oppositely charged operators with their Faraday lines extending in the same direction along the $z$ axis, the asymptotic parts of these lines then cancel, yielding an expression like \eqref{QEDline}.

However, in the gravitational case, it is easily seen that the operator $\mathcal{A}$ does not exhibit such a cancellation:
\begin{equation} \label{A}
\mathcal{A} = \phi(x) \phi(x') + V_{W_z}^\mu(x) \partial_\mu \phi(x) \phi(x') + \phi(x) V_{W_z}^\mu(x') \partial_\mu \phi(x') + O(\kappa^2).
\end{equation}
This happens because the dressing term depends on the derivative of $\phi$, rather than on $\phi$ itself as in QED.
This is a consequence of the fact that gravity couples to energy-momentum, and $\phi(x)$ is not an operator with a definite energy-momentum.

From this viewpoint the origin of the failure is obvious -- $\phi(x)\phi(x')$ is not the analog of a neutral operator in QED, since the ``charges" for gravity include momenta.  
This also suggests a resolution:  simply project this composite operator onto its zero-momentum piece.  Note that we expect to be able to accomplish this even for energy, since $\phi(x)$ contains both creation (positive energy) and annihilation (negative energy) terms.  There are two approaches to doing so.  The first is to directly introduce such a projection, by integrating over position:  let $x^{\mu\prime} = x^\mu +l^\mu$, and form the zero-momentum operator
\beq\label{bilinear}
B(l^\mu) = \int d^Dx\, \phi(x) \phi(x+l)\ ;
\eeq
one can then consider the ${\cal O}(\kappa)$ dressing of this {\it delocalized} operator.  Alternatively, we can consider the matrix element of $\Phi(x)\Phi(x')$ between two single-particle states of equal momentum.  

We first investigate this second approach, with single-particle states\footnote{Our conventions are
$\phi(x) = \int \widetilde{dp}
[a(p) e^{i p x} +  a^\dag(p) e^{-i p x} ]$,
with Lorentz-invariant measure
$\widetilde{dp} = {d^{D-1} p}/[{(2 \pi)^{D-1}2 p^0}]$.
}
 $\ket{p} = a^\dag(p) \ket{0}$ of momentum $p$; if we begin with the fully dressed operators, we have 
\begin{equation}
\bra{p} \Phi(x) \Phi(x') \ket{p} = \bra{p} \phi(x) \phi(x') \ket{p} + \left[V^\lambda(x) - V^\lambda(x') \right] \left( \braket{p|p} \bra{0} \partial_\mu \phi(x) \phi(x') \ket{0} - 2 p_\lambda \sin[ p \cdot (x - x')] \right) + O(\kappa^2).
\end{equation}
Then, we indeed see that after the zero-momentum projection, the leading-order contribution of the dressing depends on the difference between the two dressings.

Interestingly, however, this is still not sufficient to obtain a cancellation of the kind we found in QED.
This can be seen from \eqref{VWz}: the dependence of $V_{W_z,\check \mu}$ on $h_{zz}$ is linearly rising with $z$.
Thus when subtracting the two dressings the leading linear term cancels but there remains a subleading constant proportional to the separation between $x$ and $x'$.
To be concrete, let $x = (x_\perp, z)$ and $x' = (x_\perp, z')$ be two points separated along the $z$-axis, with $z' > z$.
Then we have
\begin{align}
V_{W_z,z}(x) - V_{W_z,z}(x') &= \frac{\kappa}{2} \int_z^{z'} ds\, h_{zz}(x_\perp, s), \nonumber \\
V_{W_z,\check \mu}(x) - V_{W_z,\check \mu}(x') &= \kappa \int_z^{z'} ds \; h_{z \check \mu}(x_\perp, s) + \frac{\kappa}{2} \partial_{\check \mu} \int_z^\infty ds \; \min(z' - z, s - z) h_{zz}(x_\perp, s). \label{Vdifference}
\end{align}
The function $\min(z' - z, s - z)$ rises linearly for $s$ between $z$ and $z'$, then becomes constant for $s > z'$.
Thus the first expression is entirely supported on the line between $x$ and $x'$, but the second has nontrivial dependence on the metric all the way to infinity.

In fact a little more thought makes it clear why this gravitational dressing must extend to infinity.
By considering the diagonal matrix element in the momentum basis we have restricted  to a process that creates a particle at point $x$ and annihilates one at point $x'$ (or vice versa).
This leaves the total energy and momentum unchanged, but changes the {\it angular} momentum.  Angular momentum is another  conserved Poincar\'e charge to which gravity couples.  Indeed, it may be calculated via a surface integral  -- see for example \cite{Regge1974}, and the discussion in appendix \ref{appendix:canonical}.
The nonvanishing tail of the dressing in \eqref{Vdifference} reflects the fact that the nonzero momentum of the operator must register at spatial infinity.
The lack of Poincar\'e invariance is particularly evident in the alternative approach via \eqref{bilinear}, where $l^\mu$ will transform nontrivially under a rotation or boost. 

Thus, if we want to eliminate the leading-order gravitational dressing, it seems we should project onto singlets of the full {\it Poincar\'e} group; the next section will more carefully state this requirement.  But, let us inquire whether such a nontrivial observable can be formulated by averaging $B(l^\mu)$ of \eqref{bilinear}.  We might, for example, do this by integrating over all $l^\mu$ subject to a constraint $l^2=L^2$, for some fixed distance $L$.  This, then, would be a candidate for a fully screened operator -- albeit a highly nonlocal one.

Indeed, let us consider a closely related definition of a gauge invariant bilinear operator in gravity\cite{Hamber:1984tm,Hamber:1993rb,Ambjorn:1996wc},\cite{GMH}, of a kind considered also in cosmological contexts.  Specifically, we will examine an operator of the form 
\begin{equation} \label{OO}
\mathcal{B} = \int d^D x \sqrt{|g(x)|} \; d^D x' \sqrt{|g(x')|} \; \phi(x) \phi(x') f(d(x,x'))\ ,
\end{equation}
where $d(x,x')$ is the geodesic distance between $x$ and $x'$, and $f$ is a function with support on spacelike proper distances.\footnote{Note that when working nonperturbatively, one should extend the definition to the case where there are multiple geodesics connecting $x$ and $x'$, or none at all. For our perturbative considerations, these issues do not arise.}  We can think of this operator as creating a particle, and annihilating another particle at a distance $d(x,x')$.  For example, if $f(d(x,x'))= \delta[d(x,x')-L]$, we might expect such an operator to reduce to a projected operator like we just described; the integral over both $x$ and $x'$, with $d(x,x')$ fixed, implements the projection to Poincar\'e singlets.

However, to leading order in $\kappa$, this projected operator is nearly trivial!  To see this, 
at order $\kappa^0$ we can simply insert the mode expansion for $\phi$, and replace $d(x,x')$ with the flat spacetime distance $|x - x'|$.
It is convenient to change variables to $\Delta x = x - x'$ and $\bar x = \frac12 (x + x')$, giving
\begin{equation} \label{normalO}
{:\mathcal{B}:} = \int d^D \bar x \; d^D \Delta x \; \widetilde{dp} \; \widetilde{d p'}
\Big[ (a^\dag(p) a^\dag(p') + a(p) a(p')) e^{i (p + p') \bar x + i (p - p') \tfrac12 \Delta x} + 2 a^\dag(p) a(p') e^{i (p - p') \bar x + i (p + p') \frac12 \Delta x} \Big] f(|\Delta x|).
\end{equation}

The first terms in \eqref{normalO} vanish.  
While we have dealt with an additive divergence from the last term by normal ordering, there is still an overall multiplicative divergence coming from the integral over $\bar x$.
To regulate this latter divergence in a Lorentz-invariant way we write the momentum integral as
\begin{equation} \label{intdp}
\int \widetilde{dp} = \frac{1}{(2 \pi)^{D-1}} \int d^D p \; \delta(p^2 + m^2) \theta(p^0),
\end{equation}
so that the integral over $\bar x$ in \eqref{normalO} becomes
\begin{equation}
\int d^D \bar x \; \widetilde{dp} \; \widetilde{dp'} \; e^{i (p - p') \bar x} F(p,p') = (2 \pi) \int \widetilde{dp} \; \delta(p^2 + m^2) F(p,p).
\end{equation}
The $\delta$ function in this expression  is redundant; it enforces the mass-shell condition which is already present in the integral $\widetilde{dp}$.
Our prescription for regulating this integral will be to simply drop the divergent factor $2\pi\delta$.

The resulting regulated operator is then given by
\begin{equation}  
\mathcal{B} = 2 \int d^D \Delta x \; \widetilde{dp} \; a^\dag(p) a(p) \; e^{i p \cdot \Delta x} f(|\Delta x|).
\end{equation}
The $\Delta x$ integral in this expression is manifestly  Lorentz-invariant, so it does not depend on $p$, but only on $p^2 = -m^2$.
Thus we can use this freedom to replace $p$ in the $\Delta x$ integral with $p^\mu = (m,0,0,0)$, leaving the operator
\begin{equation}
\mathcal{B} = \left( \int \widetilde{dp} \; a^\dag(p) a(p) \right) \left(  2 \int d^D \Delta x \; e^{- i m \Delta x^0} f(|\Delta x|) \right).
\end{equation}
The first factor in parentheses is the number operator, and the second is an integral that depends only on the function $f$ and the mass $m$.\footnote{For $f=\delta(d-L)$, it can be evaluated explicitly in terms of Bessel functions, but its precise form is not important for our purposes.}
So, to leading order in $\kappa$, ${\cal B}$ is simply a multiple of the number operator.  One might also ask whether there are interesting contributions at ${\cal O}(\kappa)$ or ${\cal O}(\kappa^2)$.  The former would annihilate a scalar and create another, together with a graviton, but this would require $\omega=0$ for the graviton.  The operators at ${\cal O}(\kappa^2)$ are more complicated; we leave their exploration for future work.

Indeed, the simplest way to write down nontrivial Poincar\'e singlets is instead to consider integrals of the form $\int d^Dx \sqrt{|g|} O(x)$, where $O$ is a nontrivial local scalar function of the fields and derivatives.  We will return to such observables after a more general discussion of the necessity of considering Poincar\'e singlets in order to avoid gravitational dressing.

\section{Gravitational dressing, limitations on screening, and implications for local algebras}
\label{section:dress}

In this section we prove a central result, the \emph{dressing theorem}, and discuss its implications for algebras of local observables.

\subsection{The dressing theorem}
\label{section:theorem}

The preceding discussion leads us to formulate the following theorem that constrains possible gravitational dressings.

\textbf{Dressing theorem.} 
Let $\OO$ be an operator in quantum field theory on flat spacetime, with compact support.
For example $\OO$ may be constructed out of matter fields, or out of the metric perturbation $h_{\mu \nu}$ treated as a tensor field, localized to some finite region.
Let $\tilde \OO$ be a gravitationally dressed version of $\OO$, {\it i.e.} a gauge (local diffeomorphism)-invariant operator, which has perturbative expansion about flat spacetime as  
\begin{equation}\label{opexp}
\tilde{ \mathcal{O}} = \mathcal{O} + \kappa {O}^{(1)} + \kappa^2 \mathcal{O}^{(2)} + \ldots
\end{equation}
Suppose that $\OO$ transforms nontrivially under the Poincar\'e group.
Then $\OO^{(1)}$ must depend on the asymptotic spacetime metric.

Moreover, if $\OO$ transforms nontrivially under spacetime translations, then the falloff must be
\begin{equation} \label{monopole}
\frac{\delta \OO^{(1)}}{\delta g_{\mu \nu}}  \gtrsim 1/r^{D-3}
\end{equation}
where by these falloff conditions we mean that there must exist some direction and combinations of indices $\mu$ and $\nu$ such that $\delta \OO^{(1)}/ \delta g_{\mu \nu}$ decays with radial distance $r$ no faster than $1/r^{D-3}$.
If $\OO$ transforms trivially under translations, but nontrivially under boosts or rotations, then the falloff is no faster than dipole:
\begin{equation}
\frac{\delta \OO^{(1)}}{\delta g_{\mu \nu}}  \gtrsim 1/r^{D-2}.
\end{equation}

\textbf{Proof.}  
The general idea of the proof is as follows. 
The integral of the constraint equations can be written as the sum of two terms: an integral of the stress tensor for the fields,\footnote{Including, in the case of the metric perturbation $h_{\mu\nu}$, an effective stress tensor to be defined below.} and a boundary term. 
The constraints generate gauge transformations (local diffeomorphisms), and must commute with gauge-invariant operators $\tilde {\cal O}$.
In the case of a Poincar\'e transformation, the integral of the stress tensor gives the Poincar\'e generator.
Thus if the generator does not commute with $\tilde {\cal O}$, the corresponding surface term cannot commute with $\tilde {\cal O}$.

Our proof uses the same general formalism as described in Ref.~\cite{DoGi}, where commutators of the metric perturbation $h_{\mu \nu}$ can be defined via a covariant gauge-fixed formalism.
An important property of these commutators is that they vanish outside the lightcone and hence retain locality.
While the commutators of gauge-dependent quantities such as $h_{\mu\nu}$ depend on the choice of gauge-fixing, commutators of gauge-invariant operators such as $\tilde \OO$ do not.
Thus we are free to use covariant gauge commutators in proving our result, and for these purposes it is essential that there exists a choice of gauge fixing for which the commutators are local.

Specifically, as is for example seen in Eq.~\eqref{constraint} of appendix \ref{appendix:canonical}, a local ({\it i.e.} compactly supported) diffeomorphism $\xi$ is generated by the integral of the constraint $C_\xi$ over a spatial slice $\Sigma$; 
$C_\xi$ is proportional to the Einstein equation contracted with the vector field $\xi$,
\begin{equation} \label{Cxi}
C_\xi = (-)^{D}\star\left[\left( T_{\mu\nu} - \frac{1}{8 \pi G} G_{\mu \nu} \right) \xi^\nu dx^\mu\right]\ .
\end{equation}
(Here $C_\xi$ is a $D-1$-form; $\star$ denotes Hodge dual.  For definition and conventions see appendix \ref{appendix:conventions}.)  
A gauge-invariant operator $\tilde {\cal O}$ must commute with these constraints,  
\begin{equation} \label{gconstraint}
\left[ \int_\Sigma C_\xi \, , \, \tilde \OO \right] = 0
\end{equation}
for all compactly supported $\xi$; here brackets may be regarded either classically as Poisson brackets, or as commutators in the quantum theory.
However, equation \eqref{gconstraint} must also hold for noncompact $\xi$, since we can simply write any such $\xi$ as a sum of compactly supported $\xi$.
In what follows we will take $\xi$ to be an infinitesimal isometry of Minkowski spacetime, {\it i.e.} a Poincar\'e transformation.  

We now rewrite the constraint \eqref{Cxi} by expanding the Einstein tensor perturbatively in the metric perturbation, using $g_{\mu\nu}=\eta_{\mu\nu}+\kappa h_{\mu\nu}$, as a term linear in $\kappa h_{\mu \nu}$ plus quadratic and higher-order terms,
\begin{equation} \label{Gmunu}
G_{\mu \nu} = \kappa G^{(1)}_{\mu \nu} - 8 \pi G \; t_{\mu \nu}\ ;
\end{equation}
where the quadratic and higher order terms have been absorbed into the definition of the effective gravitational stress tensor $t_{\mu \nu}$.
Substituting the expansion \eqref{Gmunu}, the constraint becomes
\begin{equation}
(\star C_\xi)_\mu = \left(-\frac{\kappa}{8 \pi G} G^{(1)}_{\mu\nu} +T_{\mu \nu} + t_{\mu \nu}   \right) \xi^\nu\ .
\end{equation}
The integral of the linear term $G^{(1)}_{\mu \nu}$ gives the Hamiltonian generator $H_\xi$ of the Poincar\'e transformation, as shown in Eq.~\eqref{Hxi}, so
\begin{equation} \label{C}
\int_\Sigma C_\xi = -H_{\xi} + \int_\Sigma \epsilon_\Sigma (T_{\mu \nu}+t_{\mu \nu}) n^\mu \xi^\nu
\end{equation}
where $\epsilon_\Sigma$ denotes the induced volume form on $\Sigma$ (here we use \eqref{sint}).
Moreover this $H_\xi$ integrates to a surface term, as shown in Eq.~\eqref{H}.
A gauge-invariant operator $\tilde{\cal O}$, satisfying \eqref{gconstraint}, therefore commutes with this surface term if and only if it commutes with the integral of $T+t$ in \eqref{C}.
If   $\tilde \OO$ has a nonzero commutator with this integral, it must depend nontrivially on the asymptotic metric.
If we expand this statement perturbatively, it becomes 
\begin{equation} \label{dressing}
\kappa [H_\xi, \mathcal{O}^{(1)}] = \left[ \int_\Sigma \epsilon_\Sigma  \, ( T_{\mu \nu}+t_{\mu \nu}) n^\mu \xi^\nu, \OO \right] + O(\kappa)\ .
\end{equation}
The left-hand side of this equation is of order $\kappa^0$, as seen in eqs.~\eqref{energy}-\eqref{boost}.\footnote{The Poincar\'e generators have a prefactor of $\kappa^{-1}$ when written in terms of the metric perturbation $h_{\mu \nu}$, or a factor $1/G$ when written in terms of the metric $g_{\mu \nu}$.} The $O(\kappa^0)$ term on the right is the commutator with the usual field theory Poincar\'e generator, including the contribution of the metric perturbation.
Thus, for $\cal O$ transforming nontrivially, the leading contribution $\calo^{(1)}$ to the dressing must depend on the asymptotic metric.

Examining the form of the generators $H_\xi$ for the different Poincar\'e transformations, Eqs.~\eqref{energy}-\eqref{boost}, we see that for spacetime translations we have 
\begin{equation}
H_\xi = \frac{1}{\kappa} \oint_{\partial \Sigma} dA\, B^{\lambda \mu \nu} \partial_\lambda h_{\mu \nu}
\end{equation}
where $B^{\lambda \mu \nu}$ is a tensor whose components are $O(r^0)$ as $r \to \infty$.
This shows that if $\delta \OO^{(1)} / \delta g_{\mu \nu}$ were to decay faster than $1/r^{D-3}$,  then the left-hand side of \eqref{dressing} would vanish, which is in contradiction with the assumed nonzero transformation law of $\OO$.  Here we use the locality of the commutators, noted above.

For boosts and rotations we have a similar expression for $H_\xi$, but now certain components of the tensor $B$ are $O(r^1)$. 
Because of the extra factor of $r$, we require only the weaker dipolelike decay rate of $1/r^{D-2}$ to achieve a nonzero value for the commutator.

\subsection{Positive energy}

The preceding argument shows that operators transforming nontrivially under the Poincar\'e group must have gravitational dressings extending to infinity.
This argument is quite general, and relies essentially only on diffeomorphism invariance, which allows the Poincar\'e charges to be expressed as boundary terms.
However, one may ask how we know that we can't cancel the contributions of $T_{\mu\nu}$ and $t_{\mu\nu}$ in \eqref{dressing} to obtain nontrivial local operators with screened Poincar\' e charges.
In classical general relativity this possibility is severely limited by the positive energy theorem.
The positive energy theorem \cite{schoen1979,witten1981} states that Minkowski spacetime is the unique global minimum of the energy, provided matter satisfies the dominant energy condition.
In conjunction with the preceding result, the positive energy theorem may place even stronger limitations on local operators.

In the context of our discussion of operators, this result can be translated as follows.  An observable can be thought of as generating a canonical transformation, mapping one solution to another, via exponentiation; this is the classical version of acting with a unitary operator such as $e^{i\lambda \OO}$, for some real $\lambda$. If we begin with the vacuum, and the transformation acts nontrivially, then by the positive energy theorem it must map to another state with a larger energy, so long as matter satisfies the dominant energy condition.  But, since the energy is a surface term, the operator therefore cannot be a local operator.  It is impossible to act purely locally on the vacuum to produce nontrivial excitations.

This discussion can be restated simply in the context of the quantum theory, if one adds the conjecture that the positive energy theorem extends to a quantum version.  
Suppose that $\OO$ is any operator that has a nontrivial matrix element between the vacuum and any other energy eigenstate $\ket{E}$, 
\begin{equation}
\bra{E} \OO \ket{0} \neq 0 \, .
\end{equation}
By positive energy, the state $\OO \ket{0}$ must have a nontrivial action under the Hamiltonian,
\begin{equation}
H \OO \ket{0} \neq 0 \, .
\end{equation}
It then follows that the operator $\OO$ must commute nontrivially with the Hamiltonian 
\begin{equation}
[H, \OO] \neq 0 \, .
\end{equation}
Since $H$ is a boundary term, any such $\OO$ must have nontrivial asymptotic support, as given by  \eqref{monopole}.

This strengthens the result of the preceding subsection, since the positive energy theorem (and, one might expect, its quantum generalization) does not refer to perturbation theory on a fixed background.
However, the positive energy considerations above do not subsume the previous result.
In particular, positive energy does not prohibit compactly supported operators $\OO$ such that $\OO \ket{0} = \alpha \ket{0}$, but which have arbitrary matrix elements among excited states.
The result of the preceding subsection goes further in constraining possible physical operators, since the spatial momentum, angular momentum, and first moment of mass (boost charge) are also boundary terms.
Hence our theorem restricts such operators $\OO$ to be Poincar\'e singlets, which is a much stronger criterion than simply preserving the vacuum.

\subsection{Discussion and other implications}

It is worth analyzing how the operators in the preceding sections are compatible with our theorem.
In the first case, we defined an operator $\mathcal{A}$  in \eqref{bidef} by simply multiplying two dressed operators \eqref{A}.
At leading order $\mathcal{A}$ reduces to the two-point operator $\phi(x) \phi(x')$.
In QED, the analogously constructed operator would define a compactly supported gauge-invariant operator provided the two operators we combine are of opposite charge and have collinear Faraday lines.
In gravity, the operator $\mathcal{A}$ is manifestly diffeomorphism-invariant, since it is constructed as a product of gauge-invariant operators. 
However, our theorem requires such an operator to have nontrivial dependence on the metric out to infinity, since at leading order the operator has nontrivial matrix elements between states with different Poincar\'{e} charges.
This explains the nontrivial gravitational tail found in Eq.~\eqref{Vdifference}: the operator $\mathcal{A}$ satisfies our theorem by failing to be compactly supported.

In the second case, we have defined an operator $\mathcal{B}$ in \eqref{OO} which is manifestly diffeomorphism-invariant.
In particular, it is also invariant under diffeomorphisms that are nontrivial asymptotically, including under the action of the Poincar\'e group.
Hence at leading order $\kappa^0$ it must have a trivial transformation law under the Poincar\'e group; and indeed we find it is simply a multiple of the number operator.
Thus the operator $\mathcal{B}$ satisfies our theorem by being a Poincar\'e singlet at leading order.

A further implication of our theorem is to establish the commutation relation between the Poincar\'{e} generators and the dressing field $V^\mu(x)$ of \eqref{dressphi}.
Specifically, it implies that for operators of the form \eqref{dressphi}, the commutation relation between the dressing field $V^\mu(x)$ and the Poincar\'{e} generator $H_\xi$ is 
\begin{equation} \label{HV}
[H_\xi, V^\mu(x)] =- i \xi^\mu(x) ,
\end{equation}
where in this equation the bracket refers to a commutator.
This result generalizes the commutation relation that was derived in Ref.~\cite{DoGi}, 
\begin{equation} \label{PV}
[P^\mu, V^\nu(x)] = i \eta^{\mu \nu},
\end{equation}
which was found by explicit calculation in some specific examples.
The result \eqref{HV} establishes that this commutation relation is a general feature of gravitational dressings, and extends \eqref{PV} to include the Lorentz generators.

\subsection{Consequences for algebra of observables}

Our dressing theorem also has direct consequences for the structure of the algebra of observables in perturbative gravity.  In short, we have shown that an  operator can only avoid dependence on the asymptotic metric, once dressed to become gauge invariant,  if it commutes with the generators of the Poincar\'e group.  This is the general statement explaining the nontrivial dressing we found in the examples of the preceding section.  
If an operator has nontrivial dependence on the asymptotic metric, it cannot be a local operator.  
Conversely, in order for an operator to commute with the momenta, it cannot be supported just on a compact region of spacetime, since the momenta would generate translations of that region. 
Therefore there are no gauge-invariant operators that can be associated with compact regions of spacetime.  
This is different from electromagnetism or non-abelian Yang Mills, where one can find local operators (such as $F_{\mu \nu}$, Wilson loops, pairs of charges connected by electric strings, {\it etc.}) that have no net charge and no dependence on the asymptotic gauge field, and which are compactly supported in spacetime.

Despite there being no local commuting algebras, one might still seek commuting subalgebras associated with regions of spacetime that extend to infinity, and so can ``contain" the gravitational field lines running off to infinity.  For example, one could dress a scalar field operator $\phi(x)$ to find an operator $\Phi_+(x)$ with a gravitational field line running to infinity parallel to the $+z$ axis, and one might expect that it would commute with a neighboring dressed operator $\Phi_-(x')$ with a field line parallel to the $-z$ axis, such that the field lines are disjoint.  However, in the full nonlinear theory there is an apparent obstacle\cite{SGalg} to defining subalgebras containing these operators.  Namely, $\Phi_+^N(x)$ can create an $N$-particle state, and thus for large $N$ can have a large energy.  In the full nonlinear theory, the support of its excitation of the gravitational field thus grows; we expect this growth to be both transverse to the $z$ direction and into the negative $z$ direction. Hence, for large enough $N$, we do not expect $\Phi_+^N(x)$ to commute with $\Phi_-^N(x)$.  
An analog of this is seen explicitly in 2+1-dimensional gravity with a negative cosmological constant \cite{DoMa}.
In this setting matter sources introduce conical deficits into spacetime, and there is a finite minimum angular support on the boundary that is determined by the mass of the source and its location in the bulk.
For two sources whose total support exceeds $2 \pi$, commuting operators cannot be defined.
This general phenomenon should be investigated more closely, but assuming it is confirmed, it even obstructs definition of commuting subalgebras, hence factorization into subsystems, associated with noncompact regions extending to asymptotic infinity.

\section{Other constructions of observables}
\label{section:other}

As indicated above, one cannot evade the theorem by finding local operators that commute with the $P_\mu$, since such operators are not localized; screening is only possible through complete delocalization.  One {\it can} however consider such nonlocal operators in order to avoid the complications of asymptotic dressing, and investigate their properties, and the question of their subalgebras and possible localization.  One example of such operators arises if one integrates a local scalar operator over all spacetime, but works in a state such that the integral is dominated by a particular region in spacetime.  Such operators have been used widely in the literature on relational observables\cite{Dewi1,Maro}\cite{GMH}.  
Another alternative is to define operators using additional, auxiliary structure such as considered in \cite{Donnelly:2016auv}. 
In either case we can ask whether subalgebra structure can be found that enables one to define locality and/or subregions, or some approximation thereof.
We turn to this question next.

\subsection{Relational field theory observables }

Operators with vanishing Poincar\'e charges can be found by integrating local operators over all of spacetime; these are the ``single integral" observables of \cite{Dewi1,Maro}\cite{GMH}.  For a general description of these, suppose that we can use the fields (including possibly the metric) to define a local $D$-form $o(x)$, which depends on the fields and their derivatives at $x$.  Then, 
\beq\label{topobs}
{\cal{O}} = \int o(x)\ ,
\eeq
with integral over the spacetime manifold, is  diffeomorphism invariant.  A particular case is that where the $D$-form arises as the dual of a scalar operator $O(x)$,
\beq\label{scdual}
o(x) = \star O(x) = \sqrt{|g|}d^Dx\, O(x) \ .
\eeq

In certain background states, such observables can be used in the semiclassical approximation to provide a localization\cite{Dewi1,Maro}\cite{GMH}.  In fact, at the semiclassical level, this is precisely how one solves the ``problem of time" in inflationary contexts:  the value of the inflaton can be used to define a time slice, and perturbations can then be defined on this slice --  {\it e.g.} on the slice where the inflaton reaches the ``reheating" value.  

A simple example is the $Z$-model of \cite{GMH}.  If, in $D$ dimensions, we have scalar fields $Z^a$, with $a=0,\ldots,D-1$, classical monotonic configurations for the $Z^a$ can be used to localize other operators.  For example, suppose we consider  a configuration where $Z^a=\lambda \delta^a_\mu x^\mu$.  We also need a bump function, which we call $f(z)$; let it vanish for $|z|>1$.  If we want to localize another observable, say $\phi(x)$, we can classically do so by defining the operator \eqref{scdual} with
\beq\label{metdep}
O_\xi(x)=\phi(x) \prod_{a=0}^{D-1} f[Z^a(x)-\xi^a]\ ,
\eeq
for any numbers $\xi^a$.  Then, clearly, the integral \eqref{topobs} will localize near the point $x^\mu = \delta^\mu_a \xi^a/\lambda$, and so classically $\cal O$ reduces to $\phi(x)$ near this point.  A ``topological" (metric-independent) observable can likewise be defined using
\beq\label{topomega}
o(x) = O_\xi(x) \prod_{a=0}^{D-1} dZ^a\ ,
\eeq
where wedge product is understood, and this observable likewise localizes.

Ref.~\cite{Maralg} argues that this classical localization can be extended to the quantum context to define localized subalgebras.  While this is an interesting topic for future investigation, there are some significant obstacles.

The first is that, even at classical level, this method only provides a localization for certain configurations of the $Z^a$ or other ``locator" fields.  For example, it is useless for providing a localization if we are discussing perturbations of empty Minkowski space.  The construction thus fails in precisely the situation where we would expect to most easily recover ordinary field theory.  In short, any such localization is {\it state dependent}, and won't be useful in a large class of states.

Even in states where this approach works, there are significant questions about how it extends to the quantum theory; some initial discussion of these issues appears in \cite{GMH,GaGi}.  Suppose that the $Z^a$ are quantum fields in a state with expectation values ({\it e.g.} resulting from appropriate boundary conditions) $\langle Z^a\rangle =\lambda \delta^a_\mu x^\mu$.  Then, we seek quantum operators like \eqref{metdep} or \eqref{topomega} with the same kind of localization properties.  The essential problem is two-fold.  First, any such operator is a rather complicated operator (due largely to the function $f$), whose quantum definition is not obvious.  Second, a quantum field $Z^a$ undergoes increasingly strong fluctuations at short distance scales, as seen from the point-split two-point function,
\beq
\langle Z(\epsilon) Z(0)\rangle \sim \frac{1}{\epsilon^{D-2}}\ ,
\eeq
with small $\epsilon$, and thus essentially takes on {\it all} values, including the desired one $\sim \xi$, in {\it any} neighborhood.  

A warm-up problem is to try to find a definition of an operator $f[Z^a(x)]$ for a single field with $\langle Z^a(x)\rangle = \lambda \delta^a_\mu x^\mu$ which approximately localizes on the codimension-one surface $x^a=0$.  Ref.~\cite{Maralg} advocates defining such an operator via power-series expansion of $f$, and using the compact support of the derivatives of $f$.  However, this is problematic, except in a formal $\hbar\rightarrow0$ limit where perturbations are taken to be infinitesimal, since such bump functions, while smooth, are nonanalytic and thus are not  reproduced by well-defined power series expansions.    Different alternatives present themselves.  One is, as in \cite{GaGi}, to instead attempt a quantum definition of $f(Z^a)$ via a Fourier transformation, and then seek a localization that is perhaps at most approximate.  Indeed, one may instead consider analytic functions of $Z^a$, such as a Gaussian, that only approximately localize as in \cite{GMH}, and investigate the departures from localization in comparison to expected Gaussian tails. 

In short, it is not clear how to use this approach to define an exact, as opposed to approximate, localization.  It is worth further investigating, systematically, the size of departures from localization, and the limits to localization, extending the analysis of \cite{GMH,GaGi}.

\subsection{Extended phase space}

Ref.~\cite{Donnelly:2016auv} proposed an approach to defining subsystems that is based on introducing, in addition to the usual metric structure of general relativity, a map $X: m \to M$ from a ``reference manifold" $m$ to the spacetime manifold $M$; $X$ plays the role of a preferred coordinatization.
One can then define a localized subsystem by choosing a region\footnote{By region we mean a region of space, {\it i.e.} an achronal spacelike codimension-1 hypersurface.} $\sigma$ of $m$, and considering its image under $X$.
The transformation law of such a map is given by
\beq\label{Xtrans}
\delta_\xi X^\mu = \xi^\mu(X)
\eeq
under diffeomorphisms, 
which is the nonlinear version of the transformation law for $V^\mu$ given in Ref.~\cite{DoGi}.
The operator
\begin{equation}\label{phiX}
\tilde \phi(x) = \phi(X(x)).
\end{equation} 
is then diffeomorphism invariant.
Ref.~\cite{Donnelly:2016auv} shows how the variables $X^\mu(x)$ can be used to define an extended phase space for gravity associated with the region $\sigma$.

To relate this construction to the preceding discussion, it is important to understand the origin and role of the $X$'s.
One possibility is that the $X$'s arise as collective coordinates for certain field degrees of freedom that can be used to define a reference system; {\it e.g.} one might think of them as coordinates of a collection of reference satellites.  In that case, their origin is in more fundamental matter degrees of freedom, and they will have a Lagrangian and stress tensor that can be inferred from those of the more basic degrees of freedom.  Then, one returns to the prior discussion:  generically the $X$s will also contribute to the energy-momentum that induces gravitational dressing; only if one defines delocalized operators with vanishing momenta does this dressing vanish.
A remaining question is of identifying the $X$'s in terms of more fundamental degrees of freedom, and of finding the stress tensor that acts on the $X$'s to generate the transformation \eqref{Xtrans}.  For example, in the $Z$ model, the $X$'s are simply the inverse functions to the $Z$ fields.

One could also ask if the existence of a physical reference frame like that described by such $X$'s is a necessary part of the structure of gravitational theories.
Suppose one accepts the conjectured AdS/CFT correspondence; interesting related work\cite{Harlow:2015lma} from this direction then provides circumstantial evidence for an affirmative answer.  Here, an extended conjecture relates bulk spacetimes with multiple asymptotic regions, such as an eternal black hole, to a Hilbert space that is a tensor product of CFTs.  This presents a puzzle, because the low-energy Hilbert space does not naturally factor in the same way when the bulk contains gauge fields (which occur whenever the CFT has global symmetries).
For example, one could consider a Wilson line stretching from one asymptotic region to the other, which does not have a factorization into low-energy gauge-invariant operators of the two asymptotic regions.  An operator that factorized would  have to look like a Wilson line to all external probes, but terminate on the horizon.  Ref.~\cite{Harlow:2015lma} suggested that this puzzle could be naturally resolved by introducing charged matter into the theory: such matter would allow the Wilson lines to end.
It was further argued that this leads to constraints on the matter content of the bulk theory, namely that these charged matter fields would automatically satisfy a form of the weak gravity conjecture.

The apparent nonfactorization of the Hilbert space occurs not only for gauge fields but for gravitational fields as well, and the latter are always present in the bulk.
One therefore encounters the same puzzle.  It may be that there is no fundamentally consistent factorization of the Hilbert space.  But if one assumes that there is, this  suggests the same kind of conjectured resolution, that 
 the gravitational analogs of Wilson lines \eqref{VWz} must be allowed to end on the bifurcation surface separating the two asymptotic regions.  And this indicates that there must be matter/fields on which all such Wilson lines can terminate.  
We have seen that gravitational Wilson lines have more constraints than their electromagnetic counterparts: while it is possible for one end of a Wilson line to end on matter (with the other extending to infinity), it does not seem to be possible to make \emph{both} ends of a Wilson line end on matter.
It is therefore an interesting question to understand the compatibility of the requirements of factorization.  Note that in the eternal AdS-Schwarzschild solution the time-translation Killing vector changes orientation at the horizon; this does suggest the possibility of a cancellation between ``positive" and ``negative" mass particles on the respective sides of the horizon, to give an unsourced Wilson line, which extends from one asymptotic region to the other.  Further investigation of constraints on matter and relation to this conjectured factorization are worth further exploration, and may even shed light on the role of  structure like the $X$'s.  Unfortunately, the argument of Ref.~\cite{Harlow:2015lma} proceeds mostly by analysis of a toy model with emergent gauge symmetry.   Analogous models of emergent gravity may be more difficult to construct and analyze; while some such models have been proposed, e.g. \cite{Gu:2009jh}, some challenges are outlined in \cite{Marolf:2014yga}.

As an alternate to deriving the $X$'s as ``collective coordinates" for matter, Ref.~\cite{Donnelly:2016auv} proposed to introduce  the $X$'s as part of the basic structure of the theory.
If the $X$'s are fundamental variables in the theory, they can be used to 
define an extended phase space associated with a region $\sigma \subset m$.
In \cite{Donnelly:2016auv} the theory carries an additional gauge symmetry, distinct from the diffeomorphism symmetry, under which $\delta g_{\mu \nu} = 0$, $\delta X^\mu = v^\mu$, where $v^\mu$ is any vector field such that $v^\mu$ and certain of its first derivatives vanish on the boundary of the region $\sigma$.
For a given $\sigma$, the $X$'s introduce an additional structure into the theory, consisting of the location of the boundary of $\sigma$ in spacetime (a spacelike codimension-2 surface without boundary), and a conformal framing of this boundary.
Defining gauge-invariant operators in this theory is not quite as straightforward as in \eqref{phiX}; while such an operator is invariant under diffeomorphisms it is \emph{not} invariant under the additional gauge symmetry that shifts $X$ in the interior.
But one can still use the surface and its framing to define observables localized in the interior, for example by shooting geodesics inward from the boundary along specified directions.
Such operators are gauge-invariant observables of the combined system consisting of the dynamical fields together with a framed surface, but not of the field theory without this additional structure.
These observables are natural generalizations of those constructed in asymptotically flat or asymptotically anti-de Sitter spacetimes where, for example, geodesics can be shot inward from the asymptotic boundary.
A primary difference is that the additional structure of the $X$'s is made explicit, whereas in the usual treatment of asymptotically flat (or asymptotically AdS) spacetimes the additional structure is implicit in the falloff conditions imposed on the metric and other dynamical fields.

Note that observables defined relative to such a subregion $\sigma$ and its boundary are {\it not} however gauge-invariant when considered from the viewpoint of a larger region in which $\sigma$ sits.
Gauge-invariant quantities in the full spacetime can be  found by combining such observables in different regions.
For example, one could first localize a surface and its framing in a gauge-invariant way, for example via the boundary of the larger region or relative to some matter fields, and construct further gauge-invariant observables relative to this surface.
Such observables must be nonlocal, or dressed, in keeping with the theorem of section \ref{section:dress}.

The introduction of the $X$'s into the fundamental theory therefore suggests a possible approach to defining a notion of subsystem different from that usually assumed in quantum field theory.
In this approach, a subsystem does not correspond to an algebra of gauge-invariant and compactly supported operators, since we have found obstructions to finding such operators.
Instead, a subsystem is  defined to consist of the values of the fields within a region together with \emph{a specification of the region under consideration};
in this sense the usual gauge invariance has been restricted.  This is essentially what the $X$'s provide: they provide both a demarcation of the subsystem, and a structure allowing one to define gauge-invariant observables of the combined system consisting of the dynamical fields and the location and framing of the boundary surface.  We leave further exploration of this possible structure for future work.

\section{Discussion and conclusion}

Local quantum field theory can be regarded as the solution to a problem, that of reconciling the principles of quantum mechanics and special relativity with the principle of locality.  In quantum field theory, the clearest formulation of the principle of locality arises in the algebraic approach, where locality is the statement that there are operator subalgebras associated with subregions of spacetime, and that subalgebras associated with spacelike-separated regions commute.  

This paper has sharpened observations of \cite{SGalg,DoGi} (with antecedents in \cite{Torre:1993fq,GMH}), that basic properties of gravity, as presently understood, obstruct such a definition of locality.  Specifically, the gauge invariance of gravity implies the inevitability of long-range gravitational fields.  We have seen that such gravitational fields must dress operators which have nonvanishing Poincar\' e charges.  Moreover, the only way that we have found to construct such Poincar\'e singlets is to average an operator over all of spacetime -- rendering it nonlocal to begin with.  The positive energy theorem reinforces this statement, at least in the classical theory, as it implies that even strong gravitational fields cannot introduce negative energy that screens the positive energy of matter.

In short, these results indicate that in a gravitational theory with weak-field behavior governed by Einstein's theory, and with positive energy, dressed operators do not give local subalgebras of the algebra of observables that reduce to those of LQFT in the small-$G$ limit.  Since such subalgebras underpin a precise definition of locality in local quantum field theory, we lack a way of describing or defining locality.  

This failure of locality extends previous discussion of related criteria for nonlocality in gravity, such as the failure of cluster decomposition pointed out in Ref. \cite{Giddings:2011xs}.
In that work it was argued that correlation functions between gravitationally dressed operators would fail to decay in the joint limit of large separation and large energy.
Whether gravitational dressing could lead to a more severe violation of cluster decomposition remains an interesting open question.

Worse than that, we also appear to lack a way to even define subsystems of a  quantum system in a gravitational theory that  reduces to a familiar definition in the weak gravity limit.
Such a concept of a subsystem, which is more basic than locality, is also an important prerequisite to discussions of aspects of quantum information, such as characterizing entanglement, entropy, and quantum information transfer. 
Indeed, while there {\it are}, as we have discussed, commuting subalgebras defining subsystems in gauge theories of internal symmetries, even in that case there are significant subtleties in combining subsystems to form a larger system.
Specifically, a proposed definition of entanglement of gauge theories in terms of subalgebras of compactly supported gauge-invariant observables \cite{Casini:2013rba} does not agree with a  Hilbert space definition of entanglement \cite{Donnelly:2014gva}, due to contributions of Wilson lines that ``cross the boundary" between the subsystems.  We have encountered even further subtleties in gravity, where there are even more challenging questions, related to the limitations on screening,  about how to treat the analogous configurations, e.g. with Wilson lines crossing a boundary.  
Whether this can be done by a prescription such as proposed in \cite{Donnelly:2016auv} remains to be seen.

One might have also expected that even if there are not local subalgebras, there could be subalgebras associated with noncompact regions, extending to infinity:  one could restrict the necessary gravitational dressing, for example, to a narrow band running off to infinity, and thus have commuting subalgebras associated with spacelike-separated such bands.  While we have not proven this is impossible, there are indicators that such a construction is forbidden, as a result of basic properties of gravity, and as is seen  in the simple case of three-dimensional spacetime.  An open question is whether gravitational theories can have any interesting commuting subalgebra structure at all, which matches onto that of LQFT in the small-$G$ limit.  Of course, if the conjectured AdS/CFT duality were  a precise equivalence (isometric isomorphism) between Hilbert spaces, that would provide a definition of commuting subalgebras associated with boundary regions; however, the relation of such a subalgebra structure with anything approximating the locality of the putative bulk spacetime remains mysterious.

We have also found obstacles to existence of certain proposed non-trivial gauge invariant operators, which are bilinears in field operators, and have been considered in the cosmological context (see \cite{GiSl} and references therein).  In the case of perturbation theory about a Minkowski-like state, such operators are, at leading order in the gravitational coupling $\kappa$, nearly trivial, specifically reducing to the number operator.  

Thus, the most promising kind of gauge-invariant observables to consider are the ``single-integral" observables\cite{Dewi1,Maro}\cite{GMH}, which were discussed in section \ref{section:other}.  
In certain states, these can provide an approximate notion of locality, by allowing localization with respect to features of the state, as has been for example discussed in \cite{GMH,GaGi,Maralg}.  
However, as was also pointed out in \cite{GMH}, and is seen from a different direction here, such a definition of locality is only approximate, and moreover is state-dependent.  
 The weak-field behavior of gravity, which we expect to be recovered from any more basic theory, this seems to indicate that locality in general is both approximate and state-dependent; it remains to be seen if this can be circumvented in some more fundamental approach.

\section{Acknowledgements}

We wish to thank N. Afshordi, B. Dittrich, J. Hartle, I. Khavkine, and C. Rovelli,  for helpful discussions, and especially D. Marolf for helpful discussions and for comments on a draft of this work.
We also thank an anonymous referee for helpful comments.
This work was supported in part by the U.S. DOE under Contract No. 
DE-SC0011702,  and by Foundational Questions Institute (fqxi.org) 
Grant No. FQXi-RFP-1507.

\appendix
\section{Some conventions}
\label{appendix:conventions}
In this appendix we briefly summarize some of the conventions used in the paper.

A $p$-form $B$ is related to its components  by
\beq
B= \frac{1}{p!} B_{\mu_1 \cdots\mu_p} dx^{\mu_1} \cdots dx^{\mu_p}
\eeq
where wedge product is understood between differentials $dx^\mu$.  
In $D=d+1$ dimensional spacetime, we define the completely antisymmetric Levi-Civita tensor
\beq
\epsilon_{01\cdots d}= \sqrt{|g|}\ ,
\eeq
with other permutations differing by the sign of the permutation.
The volume form is defined as
\beq
\epsilon=\frac{1}{D!} \epsilon_{\mu_1\cdots\mu_D}dx^{\mu_1}\cdots dx^{\mu_D}\ .
\eeq

The Hodge dual of a $p$-form is defined as 
\beq
(\star B)_{\mu_1\cdots\mu_{D-p}}=\frac{1}{p!}\epsilon_{\mu_1\cdots\mu_{D-p}}{}^{\nu_1\cdots\nu_p} B_{\nu_1\cdots\nu_p}\ ;
\eeq
note that this common convention
differs from Wald's convention \cite{Wald:1984rg} by a factor of $(-1)^{p(D-p)}$ acting on $p$-forms.  
Repeated dual gives $\star\star=(-1)^{p(D-p)+1}$ in Lorentzian signature, so $\star^2 = (-1)^D$ acting on one-forms, and $\star^2 = -1$ acting on two-forms.

The integral of a $(D-1)$-form over a  $(D-1)$-dimensional spatial surface $\Sigma$ is given in terms of its dual components by
\begin{equation} \label{sint}
\int_\Sigma A = \int_\Sigma \epsilon_\Sigma (\star A)_\mu n^\mu
\end{equation}
where $\epsilon_\Sigma$ is the induced volume form on $\Sigma$, and $n^\mu$ is the unit normal.

\section{The covariant canonical formalism}
\label{appendix:canonical}

In this appendix, we give a brief overview of the covariant canonical formalism for gravity, which is used in the proof of section \ref{section:dress}.
This formalism has been developed in a large number of works, for example \cite{ashtekar1982, Ashtekar:1987hia,Crnkovic:1986ex,Crnkovic:1987tz,Zuckerman:1989cx,Lee:1990nz,Iyer:1994ys}.
An advantage of this formalism for our purposes is that it identifies the phase space directly with the space of physical solutions, allowing us to consider Poisson brackets between observables defined at different times, rather than having to express all such  observables in terms of initial data on a single time slice \cite{Marolf1993a}.

The space of solutions for general relativity can be endowed with a conserved symplectic structure as follows.
We start with a Lagrangian $D$-form $L$, which describes general relativity coupled to a scalar field,
\begin{equation}
L = \frac{1}{16 \pi G} R \epsilon - \left[ \frac12 \nabla_\mu \phi \nabla^\mu \phi + V(\phi) \right]\epsilon\ .
\end{equation}
Here $G$ is the $D$-dimensional Newton constant and $\epsilon$ denotes the volume form (for conventions, see Appendix \ref{appendix:conventions}); we also often substitute $\kappa^2=32\pi G$.  

This Lagrangian is usually supplemented with a boundary term \cite{York,Regge1974,Hawking:1995fd,MaMa}
that is necessary to have a good variational principle subject to fixed asymptotic conditions, \emph{i.e.} that the action be stationary under all variations preserving the asymptotic conditions, given the local equations of motion.
While we could include such a boundary term at this stage,  we will see that it is not needed for our discussion of the symplectic structure and derivation of the Poincar\'e generators.  
In the covariant canonical formalism, the appropriate boundary terms for the Hamiltonian can also be fixed directly from the falloff conditions as in 
\cite{Iyer:1994ys} and as we will describe, rather than derived from the boundary term in the Lagrangian as in \cite{Hawking:1995fd}.

The symplectic potential $(D-1)$-form $\theta$ is defined by varying the Lagrangian,\footnote{Here we use the convention that $\delta g^{\mu\nu}$ is the variation of the inverse metric.}
\begin{equation} \label{deltaL}
\delta L = E_{\mu \nu} \delta g^{\mu \nu} + E \delta \phi + d \theta.
\end{equation}
The densities $E_{\mu \nu}$ and $E$ are proportional to the equations of motion, which are the Einstein equation and the scalar field equation respectively:
\begin{equation}
E_{\mu \nu} = \left( \frac{1}{16 \pi G} G_{\mu \nu} - \frac{1}{2} T_{\mu \nu}  \right) \epsilon\quad , \qquad E = [\nabla^2 \phi - V'(\phi)] \epsilon\ .
\end{equation}
$\theta$ is the symplectic potential $(D-1)$-form, which depends linearly on the variations $\delta g_{\mu \nu}$ and $\delta \phi$:
\begin{equation}\label{symppot}
(\star \theta)_\mu = -\frac{1}{16 \pi G} (\nabla^\nu \delta g_{\mu \nu} - \nabla_\mu \delta g)
+ (\nabla_\mu \phi) \delta \phi,
\end{equation}
here written in terms of its dual via the Hodge $\star$ operator.  

One can think of an expression such as \eqref{symppot} as giving a differential form on {\it field} space, where e.g. $\delta g_{\mu\nu}(x)$, $\delta\phi(x)$ give a basis for one-forms.  The variational derivative $\delta$ can be thought of as an exterior derivative on this field space, so raises the degree of such a form by one and satisfies $\delta^2=0$.  

The presymplectic form $\omega =\delta \theta$ is then obtained by varying $\theta$.
Thus $\omega$ is a $(D-1)$-form in spacetime and a 2-form in field space.
Varying \eqref{deltaL} we see that $\omega$ is closed, $d\omega=0$, whenever its arguments are linearized solutions. 
It then follows that the symplectic form 
\begin{equation} \label{Omega}
\Omega = \int_\Sigma \omega
\end{equation}
is independent of the Cauchy surface $\Sigma$. 
More precisely the symplectic form on two Cauchy surfaces $\Sigma_1$ and $\Sigma_2$ are equal on-shell if $\partial \Sigma_1 = \partial \Sigma_2$. 
In general one might wish to consider two surfaces $\Sigma_1$ and $\Sigma_2$ separated in time; in that case the symplectic form is preserved only provided the boundary conditions are sufficiently strong that no symplectic flux can escape between the two slices.
The space of solutions to the equations of motion with the form $\Omega$ defines a presymplectic manifold: $\Omega$ is closed but is degenerate.  
In this approach the phase space corresponds to the space of solutions.

In order to have a well-defined phase space, the symplectic form \eqref{Omega} should be finite.
If $\Sigma$ consists of a spacelike slice of asymptotically flat spacetime, this means imposing suitable falloff conditions on the metric.
We will adopt standard asymptotically flat conditions such that $g_{\mu \nu} = \eta_{\mu \nu} + \kappa h_{\mu \nu}$, with $h_{\mu \nu}$ satisfying the falloff conditions
\begin{equation} \label{falloff}
h_{\mu \nu} = O(1/r^{D-3}), \qquad \partial_\lambda h_{\mu \nu} = O(1/r^{D-2}).
\end{equation} 
For $D > 4$, these falloff conditions are sufficient to ensure a finite symplectic structure.
However, for $D=4$ the falloff conditions alone are insufficient.
The symplectic form can be written 
\begin{equation}
\Omega = \int_\Sigma \epsilon_{\Sigma}\, T^{\alpha \beta \lambda \mu \nu}{}_\sigma {\delta} g_{\alpha \beta} \partial_\lambda {\delta} g_{\mu \nu} n^\sigma
\end{equation}
where $\epsilon_{\Sigma}$ is the induced volume form on 
$\Sigma$, $n^\sigma$ the unit normal, and $T$ is a tensor constructed from the metric, see e.g. \cite{Hollands:2012sf}.
The falloff conditions ensure that the integrand is $O(1/r^3)$, which is not sufficient to rule out a logarithmic divergence in $\Omega$.
To avoid this potential divergence, one can impose the parity condition \cite{Regge1974}: that the $O(1/r)$ piece of $h_{\mu \nu}$ is \emph{even} under parity ($x^\mu \to -x^\mu$), whereas the $O(1/r^2)$ piece of $\partial_\lambda h_{\mu \nu}$ is parity \emph{odd}.
This condition is satisfied by the Schwarzschild solution in its standard coordinates, and is preserved under the action of the Poincar\'e group.
Thus this phase space is sufficient to describe any number of particles accompanied by their gravitational fields.\footnote{For some work attempting to relax this parity condition see Ref.~\cite{Compere:2011ve}.}
We will also assume falloff conditions on the matter field $\phi$ so that its contribution to $\Omega$ is finite.
These conditions would have to be relaxed further to account for massless fields such as the electromagnetic field, but we will not consider this generalization here.

The symplectic form allows us to associate conserved currents and charges that act as Hamiltonian generators with symmetries of the phase space.
Let $\xi$ be a vector field in spacetime, specifying a diffeomorphism. 
We  associate with $\xi$ a vector field in field space where all the fields are transformed by minus their Lie derivatives along $\xi$.
Let ${\cal I}_\xi$ denote the interior product with this vector field in field space, defined by
\begin{equation}
{\cal I}_\xi \delta g_{\mu \nu} = - \mathcal{L}_\xi g_{\mu \nu} = -\nabla_\mu \xi_\nu - \nabla_\nu \xi_\mu, \qquad {\cal I}_\xi \delta \phi = -\mathcal{L}_\xi \phi = -\xi^\mu \nabla_\mu \phi.
\end{equation}
We can then define the Noether current $(D-1)$-form:
\beq\label{notherdef}
J_\xi = -{\cal I}_\xi \theta - i_\xi L \,
\eeq
which becomes
\beq
(\star J_\xi)_\mu = -\frac{1}{16 \pi G} \nabla^\nu (\nabla_\nu \xi_\mu - \nabla_\mu \xi_\nu) + \left( T_{\mu\nu} - \frac{1}{8 \pi G} G_{\mu \nu} \right) \xi^\nu\ .
\eeq
The current $J_\xi$ is conserved on shell, which can be seen  directly by writing it in the form
\begin{equation} \label{noethercharge}
J_\xi = d Q_\xi + C_\xi
\end{equation}
where $Q_\xi$ is the Noether charge $(D-2)$-form \cite{Wald:1993nt,Iyer:1994ys}, and $C_\xi$ is the constraint, which vanishes on shell:
\begin{equation} \label{QC}
(\star Q_\xi)_{\mu \nu} = \frac{1}{16 \pi G} (\nabla_\mu \xi_\nu - \nabla_\nu \xi_\mu)= \frac{1}{16 \pi G}(d\xi)_{\mu\nu}, \qquad (\star C_\xi)_\mu = \left( T_{\mu\nu} - \frac{1}{8 \pi G} G_{\mu \nu} \right) \xi^\nu.
\end{equation}
Here $\xi_\mu$ is the one-form obtained by lowering the index on $\xi^\mu$.

To find the canonical generator $H_\xi$ of an infinitesimal diffeomorphism $\xi$, we need to find a functional that generates the transformation $\delta_\xi = - \mathcal{L}_\xi$ on all dynamical fields.
Hence its variation must solve the equation
\begin{equation} \label{deltaH}
\delta H_\xi = {\cal I}_\xi \Omega.
\end{equation}
This can be understood by considering a finite-dimensional phase space with symplectic form $\omega_{ab}$.
When $\omega$ is invertible, the Poisson brackets are defined by $\{ f, g \} =  (\omega^{-1})^{ab} \partial_a f \partial_b g$. 
Then \eqref{deltaH} is equivalent to the familiar statement that the Hamiltonian $H_\xi$ generates the phase space flow $\xi$ via the Poisson brackets:
\begin{equation}
\delta_\xi f = \{ H_\xi,f\}\ ,
\end{equation}
where $f$ is a function of phase-space variables.
An advantage an approach based on \eqref{deltaH} is that it does not require $\Omega$ to be inverted.

We find such a functional by first finding ${\cal I}_\xi \Omega$.  To do this, begin with the variation of the Noether expression \eqref{notherdef},
which may be rewritten using the relation
\beq
\delta {\cal I}_\xi \theta + {\cal I}_\xi \delta\theta = -\mathcal{L}_\xi\theta
\eeq
and the variation \eqref{deltaL} to give
\begin{equation} \label{deltaJ}
\delta J_\xi =  {\cal I}_\xi \omega - i_\xi E_{\mu \nu} \delta g^{\mu \nu} - i_\xi E \delta \phi + d (i_\xi \theta).
\end{equation}
Substituting \eqref{noethercharge} for $J_\xi$ and integrating over a slice $\Sigma$, this implies the identity
\begin{equation} \label{onshell}
{\cal I}_\xi \int_\Sigma \omega = \delta \int_\Sigma C_\xi +  \oint_{\partial \Sigma}\left( \delta Q_\xi -i_\xi \theta\right) + \int_\Sigma i_\xi \left( E_{\mu\nu} \delta g^{\mu\nu} + E \delta \phi\right )\ .
\end{equation}
On-shell, the terms proportional to $E^{\mu \nu}$ and $E$ vanish.  
Using this to solve the  equation \eqref{deltaH} for $H_\xi$ allows us to find the generator for a given infinitesimal diffeomorphism.

We will be interested in two types of transformations.
First, consider the case where $\xi$ is compactly supported on the interior of $\Sigma$. 
In that case the boundary terms vanish; for generic perturbations about a solution, the terms $E^{\mu\nu}$ and $E$ also vanish.  Formally,
the generator is simply the integral of the constraint: 
\begin{equation} \label{constraint}
H_\xi = \int_\Sigma C_\xi,  \qquad \text{for } \xi \text{ compactly supported}.
\end{equation}
This generator vanishes on shell; the corresponding diffeomorphism is  a gauge symmetry.

The second type of transformation we will be interested in is the Poincar\'{e} symmetries.
In this case $\xi^\mu$ is not compactly supported, so the generator is not just proportional to the constraints.
Instead $H_\xi$ can be written on shell as
\beq
H_\xi = \oint_{\partial \Sigma} \left( Q_\xi + B_\xi \right)\ , 
\eeq
where $B_\xi$ is an additional boundary term satisfying 
\beq\label{Bdef}
\delta \oint_{\partial \Sigma} B_\xi = -\oint_{\partial \Sigma} i_\xi \theta\ .
\eeq
In order for  $H_\xi$ to exist, asymptotic conditions must be imposed on the metric.  In particular, by \eqref{deltaH}, $\delta {\cal I}_\xi \Omega=0$;  taking $\delta$ of \eqref{onshell} shows that we require
\beq \label{zeroflux}
\oint_{\partial\Sigma} i_\xi \omega = 0\ .
\eeq
From the falloff conditions \eqref{falloff} we see that this condition is satisfied whenever $\xi = O(r^\alpha)$ with $\alpha < D - 3$ as $r \to \infty$.
This includes the translation subgroup of the Poincar\'{e} group for $D > 3$.
In the case of a Lorentz generator, the corresponding vector field $\xi^\mu$ is $O(r^1)$, so the zero flux condition is guaranteed by the falloff conditions for $D > 4$.
However, $\xi^\mu$ in this case is parity odd, and so in $D = 4$ the parity condition on the metric ensures that the zero flux condition \eqref{zeroflux} is satisfied.
Thus the Poincar\'e generators are well defined for all $D \geq 4$.

We can solve for $B_\xi$ by simply replacing the occurrences of $\delta g$ in $\theta$ with the linearized perturbation $\kappa h_{\mu \nu}$.
Denoting $I_{\kappa h} \delta g_{\mu \nu} = \kappa h_{\mu \nu}$, we can write the generators as
\begin{equation} \label{H}
H_\xi =\oint_{\partial\Sigma} (Q_\xi -I_{\kappa h} i_\xi \theta)= I_{\kappa h} \oint_{\partial \Sigma} (\delta Q_\xi - i_\xi \theta).
\end{equation}

In the case where $\xi$ is a Poincar\'e symmetry, we can also express the generator as an integral of the linearized Einstein tensor.
Around a background with flat metric and vanishing fields,  ${\cal I}_\xi$ of any form is zero, and the background equations of motion are satisfied.
In that case \eqref{onshell} yields
\begin{equation}
-\int_\Sigma \delta C_\xi = \oint_{\partial \Sigma} (\delta Q_\xi - i_\xi \theta).
\end{equation}
Contracting with $I_{\kappa h}$, we see that
\begin{equation}
-I_{\kappa h} \int_\Sigma \delta C_\xi = H_\xi.
\end{equation}
Using the form of the constraint \eqref{QC}, we then see that the left-hand side is simply the linearized Einstein tensor, integrated over a slice:
\begin{equation} \label{Hxi}
H_\xi =  \frac{1}{8 \pi G} I_{\kappa h} \delta \int_\Sigma \epsilon_\Sigma  \; n^\mu G_{\mu\nu} \xi^\nu,
\end{equation}
where, as above, $I_{\kappa h} \delta$ just means to expand the quantity to linear order in the metric perturbation.
If we explicitly linearize about flat space and split this into timelike and spacelike components, we obtain
\begin{equation} \label{linG}
H_{\xi} = \frac{2}{\kappa} \int d^{D-1} x \Big[ \xi^0 (\partial_i \partial_j h_{ij} - \partial_i^2  h_{jj}) + \xi^i \partial_j (\partial_0 h_{ij} - \delta_{ij} \partial_0 h_{kk}  + {\partial_i h_{0j}} - \partial_j h_{0i}) \Big].
\end{equation}

Finally, we will make use of the explicit form of the Poincar\'e generators.
These can be obtained either by integrating the linearized Einstein tensor in \eqref{linG}, or by solving \eqref{H}.
The result is 
\begin{align}
\label{energy}P^0 &= \frac{2}{\kappa}\oint_{\partial\Sigma} dA\, \hat r^i (\partial_j h_{ij} - \partial_i h_{jj} ), \\
P^i &= - \frac{2}{\kappa} \oint_{\partial\Sigma} dA\, \hat r^j (\partial_0 h_{ij} - \delta_{ij}\partial_0 h_{kk}  + \partial_i h_{0j} - \partial_j h_{0i}), \\
L^{ij} &=-\frac{2}{\kappa} \oint_{\partial\Sigma} dA\, \hat r^k \left[ x^i (\partial_0 h_{jk} - \partial_k h_{0j}) + h_{0j} \delta_{ik} \right] - (i\leftrightarrow j),  \\
\label{boost}K^i &=  \frac{2}{\kappa} \oint_{\partial\Sigma} dA\, \hat r^j \left[ x^i (\partial_k h_{jk} - \partial_j h_{kk}) - h_{ij} + h_{kk} \delta_{ij} + x^0 (\partial_0 h_{ij} - \delta_{ij} \partial_0 h_{kk}  + \partial_i h_{0j} 
- \partial_j h_{0i} )\right].
\end{align}
These generate transformations  $\xi = \partial_0$, $\xi = -\partial_i$, $\xi =  -x^i \partial_j + x^j \partial_i$, and $\xi = x^i \partial_0 + x^0 \partial_i$, respectively. 
It can be verified that these agree with the standard expressions, given e.g. in Ref.~\cite{Regge1974}.

\bibliographystyle{utphys}
\bibliography{QG-obs}

\providecommand{\href}[2]{#2}\begingroup\raggedright\begin{thebibliography}{10}

\bibitem{Haag}
R.~Haag, {\em {Local quantum physics: Fields, particles, algebras}}.
\newblock (Texts and monographs in physics). Springer, Berlin, Germany,
1992.
\newblock

\bibitem{SGalg}
S.~B. Giddings, ``{Hilbert space structure in quantum gravity: an algebraic
  perspective},'' \href{http://dx.doi.org/10.1007/JHEP12(2015)099}{{\em JHEP}
  {\bfseries 12} (2015) 099},
\href{http://arxiv.org/abs/1503.08207}{{\ttfamily arXiv:1503.08207 [hep-th]}}.

\bibitem{DoGi}
W.~Donnelly and S.~B. Giddings, ``{Diffeomorphism-invariant observables and
  their nonlocal algebra},''
  \href{http://dx.doi.org/10.1103/PhysRevD.93.024030}{{\em Phys. Rev.}
  {\bfseries D93} no.~2, (2016) 024030},
  \href{http://arxiv.org/abs/1507.07921}{{\ttfamily arXiv:1507.07921
  [hep-th]}}.
[Erratum: Phys. Rev.D94,no.2,029903(2016)].

\bibitem{Torre:1993fq}
C.~G. Torre, ``{Gravitational observables and local symmetries},''
  \href{http://dx.doi.org/10.1103/PhysRevD.48.R2373}{{\em Phys. Rev.}
  {\bfseries D48} (1993) R2373--R2376},
\href{http://arxiv.org/abs/gr-qc/9306030}{{\ttfamily arXiv:gr-qc/9306030
  [gr-qc]}}.

\bibitem{GMH}
S.~B. Giddings, D.~Marolf, and J.~B. Hartle, ``{Observables in effective
  gravity},'' \href{http://dx.doi.org/10.1103/PhysRevD.74.064018}{{\em
  Phys.Rev.} {\bfseries D74} (2006) 064018},
\href{http://arxiv.org/abs/hep-th/0512200}{{\ttfamily arXiv:hep-th/0512200
  [hep-th]}}.

\bibitem{Dewi1}
B.~S. DeWitt, ``{The Quantization of geometry},'' in {\em Gravitation: An
  Introduction to Current Research}, L.~Witten, ed., pp.~266--381.
\newblock Wiley, New York,
1962.
\newblock

\bibitem{Maro}
D.~Marolf, ``{Quantum observables and recollapsing dynamics},''
  \href{http://dx.doi.org/10.1088/0264-9381/12/5/011}{{\em Class.Quant.Grav.}
  {\bfseries 12} (1995) 1199--1220},
\href{http://arxiv.org/abs/gr-qc/9404053}{{\ttfamily arXiv:gr-qc/9404053
  [gr-qc]}}.

\bibitem{GeDe1}
J.~G{\'e}h{\'e}niau and R.~Debever, ``Les invariants de courbure de l'espace de
  Riemann a quatre dimensions,'' {\em Bull. Cl. Sci. Acad. R. Belg} {\bfseries
  42} (1956) 114.

\bibitem{GeDe2}
J.~G{\'e}h{\'e}niau and R.~Debever, ``Les quatorze invariants de courbure de
  l'espace Riemannien a quatre dimensions,'' {\em Helv. Phys. Acta Suppl}
  {\bfseries 4} (1956) 101--105.

\bibitem{Komar}
A.~Komar, ``Construction of a complete set of independent observables in the
  general theory of relativity,'' {\em Phys. Rev.} {\bfseries 111} (1958) 1182.

\bibitem{BeKo}
P.~G. Bergmann and A.~B. Komar, ``Poisson Brackets Between Locally Defined
  Observables in General Relativity,''
  \href{http://dx.doi.org/10.1103/PhysRevLett.4.432}{{\em Phys. Rev. Lett.}
  {\bfseries 4} (Apr, 1960) 432--433}.

\bibitem{Donnelly:2016auv}
W.~Donnelly and L.~Freidel, ``{Local subsystems in gauge theory and gravity},''
  \href{http://dx.doi.org/10.1007/JHEP09(2016)102}{{\em JHEP} {\bfseries 09}
  (2016) 102},
\href{http://arxiv.org/abs/1601.04744}{{\ttfamily arXiv:1601.04744 [hep-th]}}.

\bibitem{Dirac1955}
P.~A. Dirac, ``{Gauge-invariant formulation of quantum electrodynamics},''
\href{http://dx.doi.org/10.1139/p55-081}{{\em Can.J.Phys.} {\bfseries 33}
  (1955) 650}.

\bibitem{GAK}
S.~Giddings, J.~Abbott, and K.~Kuchar, ``{Einstein's theory in a
  three-dimensional space-time},''
\href{http://dx.doi.org/10.1007/BF00762914}{{\em Gen. Rel. Grav.} {\bfseries
  16} (1984) 751--775}.

\bibitem{Regge1974}
T.~Regge and C.~Teitelboim, ``Role of surface integrals in the Hamiltonian
  formulation of general relativity,''
  \href{http://dx.doi.org/10.1016/0003-4916(74)90404-7}{{\em Annals of Physics}
  {\bfseries 88} no.~1, (1974) 286 -- 318}.

\bibitem{Hamber:1984tm}
H.~W. Hamber, ``{Simplicial quantum gravity},'' in {\em {Summer School in
  Theoretical Physics, Session XLIII: Critical Phenomena, Random Systems, Gauge
  Theories Les Houches, France, August 1-September 7, 1984}}.
\newblock
1984.
\newblock

\bibitem{Hamber:1993rb}
H.~W. Hamber, ``{Invariant correlations in simplicial gravity},''
  \href{http://dx.doi.org/10.1103/PhysRevD.50.3932}{{\em Phys. Rev.} {\bfseries
  D50} (1994) 3932--3941},
\href{http://arxiv.org/abs/hep-th/9311024}{{\ttfamily arXiv:hep-th/9311024
  [hep-th]}}.

\bibitem{Ambjorn:1996wc}
J.~Ambjorn and K.~N. Anagnostopoulos, ``{Quantum geometry of 2D gravity coupled
  to unitary matter},''
  \href{http://dx.doi.org/10.1016/S0550-3213(97)00259-9}{{\em Nucl. Phys.}
  {\bfseries B497} (1997) 445--478},
\href{http://arxiv.org/abs/hep-lat/9701006}{{\ttfamily arXiv:hep-lat/9701006
  [hep-lat]}}.

\bibitem{schoen1979}
R.~Schoen and S.~T. Yau, ``On the proof of the positive mass conjecture in
  general relativity,'' \href{http://dx.doi.org/10.1007/BF01940959}{{\em Comm.
  Math. Phys.} {\bfseries 65} no.~1, (1979) 45--76}.

\bibitem{witten1981}
E.~Witten, ``A new proof of the positive energy theorem,''
  \href{http://dx.doi.org/10.1007/BF01208277}{{\em Comm. Math. Phys.}
  {\bfseries 80} no.~3, (1981) 381--402}.

\bibitem{DoMa}
W.~Donnelly, D.~Marolf, and E.~Mintun, ``{Combing gravitational hair in 2 + 1
  dimensions},'' \href{http://dx.doi.org/10.1088/0264-9381/33/2/025010}{{\em
  Class. Quant. Grav.} {\bfseries 33} no.~2, (2016) 025010},
\href{http://arxiv.org/abs/1510.00672}{{\ttfamily arXiv:1510.00672 [hep-th]}}.

\bibitem{Maralg}
D.~Marolf, ``{Comments on Microcausality, Chaos, and Gravitational
  Observables},'' \href{http://dx.doi.org/10.1088/0264-9381/32/24/245003}{{\em
  Class. Quant. Grav.} {\bfseries 32} no.~24, (2015) 245003},
\href{http://arxiv.org/abs/1508.00939}{{\ttfamily arXiv:1508.00939 [gr-qc]}}.

\bibitem{GaGi}
M.~Gary and S.~B. Giddings, ``{Relational observables in 2D quantum gravity},''
  \href{http://dx.doi.org/10.1103/PhysRevD.75.104007}{{\em Phys.Rev.}
  {\bfseries D75} (2007) 104007},
\href{http://arxiv.org/abs/hep-th/0612191}{{\ttfamily arXiv:hep-th/0612191
  [hep-th]}}.

\bibitem{Harlow:2015lma}
D.~Harlow, ``{Wormholes, Emergent Gauge Fields, and the Weak Gravity
  Conjecture},'' \href{http://dx.doi.org/10.1007/JHEP01(2016)122}{{\em JHEP}
  {\bfseries 01} (2016) 122},
\href{http://arxiv.org/abs/1510.07911}{{\ttfamily arXiv:1510.07911 [hep-th]}}.

\bibitem{Gu:2009jh}
Z.-C. Gu and X.-G. Wen, ``{Emergence of helicity $\pm$ 2 modes (gravitons) from
  qubit models},''
  \href{http://dx.doi.org/10.1016/j.nuclphysb.2012.05.010}{{\em Nucl. Phys.}
  {\bfseries B863} (2012) 90--129},
\href{http://arxiv.org/abs/0907.1203}{{\ttfamily arXiv:0907.1203 [gr-qc]}}.

\bibitem{Marolf:2014yga}
D.~Marolf, ``{Emergent Gravity Requires Kinematic Nonlocality},''
  \href{http://dx.doi.org/10.1103/PhysRevLett.114.031104}{{\em Phys. Rev.
  Lett.} {\bfseries 114} no.~3, (2015) 031104},
\href{http://arxiv.org/abs/1409.2509}{{\ttfamily arXiv:1409.2509 [hep-th]}}.

\bibitem{Giddings:2011xs}
S.~B. Giddings, ``{The gravitational S-matrix: Erice lectures},''
  \href{http://dx.doi.org/10.1142/9789814522489_0005}{{\em Subnucl. Ser.}
  {\bfseries 48} (2013) 93--147},
\href{http://arxiv.org/abs/1105.2036}{{\ttfamily arXiv:1105.2036 [hep-th]}}.

\bibitem{Casini:2013rba}
H.~Casini, M.~Huerta, and J.~A. Rosabal, ``{Remarks on entanglement entropy for
  gauge fields},'' \href{http://dx.doi.org/10.1103/PhysRevD.89.085012}{{\em
  Phys. Rev.} {\bfseries D89} no.~8, (2014) 085012},
\href{http://arxiv.org/abs/1312.1183}{{\ttfamily arXiv:1312.1183 [hep-th]}}.

\bibitem{Donnelly:2014gva}
W.~Donnelly, ``{Entanglement entropy and nonabelian gauge symmetry},''
  \href{http://dx.doi.org/10.1088/0264-9381/31/21/214003}{{\em Class. Quant.
  Grav.} {\bfseries 31} no.~21, (2014) 214003},
\href{http://arxiv.org/abs/1406.7304}{{\ttfamily arXiv:1406.7304 [hep-th]}}.

\bibitem{GiSl}
S.~B. Giddings and M.~S. Sloth, ``{Fluctuating geometries, q-observables, and
  infrared growth in inflationary spacetimes},''
  \href{http://dx.doi.org/10.1103/PhysRevD.86.083538}{{\em Phys.Rev.}
  {\bfseries D86} (2012) 083538},
\href{http://arxiv.org/abs/1109.1000}{{\ttfamily arXiv:1109.1000 [hep-th]}}.

\bibitem{Wald:1984rg}
R.~M. Wald,
  \href{http://dx.doi.org/10.7208/chicago/9780226870373.001.0001}{{\em {General
  Relativity}}}.
\newblock University of Chicago Press, 1984.

\bibitem{ashtekar1982}
A.~Ashtekar and A.~Magnon-Ashtekar, ``On the symplectic structure of general
  relativity,'' {\em Comm. Math. Phys.} {\bfseries 86} no.~1, (1982) 55--68.

\bibitem{Ashtekar:1987hia}
A.~Ashtekar, L.~Bombelli, and R.~Koul, ``{Phase space formulation of general
  relativity without a 3+1 splitting},''
\href{http://dx.doi.org/10.1007/3-540-17894-5_378}{{\em Lect. Notes Phys.}
  {\bfseries 278} (1987) 356--359}.

\bibitem{Crnkovic:1986ex}
C.~Crnkovic and E.~Witten, ``{Covariant description of canonical formalism in
  geometrical theories},'' in {\em Three hundred years of gravitation},
  S.~Hawking and W.~Israel, eds.
\newblock
1986.
\newblock

\bibitem{Crnkovic:1987tz}
C.~Crnkovic, ``{Symplectic Geometry of the Covariant Phase Space},''
\href{http://dx.doi.org/10.1088/0264-9381/5/12/008}{{\em Class. Quant. Grav.}
  {\bfseries 5} (1988) 1557--1575}.

\bibitem{Zuckerman:1989cx}
G.~J. Zuckerman, ``{Action principles and global geometry},'' in {\em
  Mathematical Aspects of String Theory: proceedings}, S.~Yau, ed.,
  pp.~259--284.
\newblock World Scientific,
1987.
\newblock

\bibitem{Lee:1990nz}
J.~Lee and R.~M. Wald, ``{Local symmetries and constraints},''
\href{http://dx.doi.org/10.1063/1.528801}{{\em J. Math. Phys.} {\bfseries 31}
  (1990) 725--743}.

\bibitem{Iyer:1994ys}
V.~Iyer and R.~M. Wald, ``{Some properties of Noether charge and a proposal for
  dynamical black hole entropy},''
  \href{http://dx.doi.org/10.1103/PhysRevD.50.846}{{\em Phys. Rev.} {\bfseries
  D50} (1994) 846--864},
\href{http://arxiv.org/abs/gr-qc/9403028}{{\ttfamily arXiv:gr-qc/9403028
  [gr-qc]}}.

\bibitem{Marolf1993a}
D.~M. Marolf, ``{Poisson brackets on the space of histories},''
  \href{http://dx.doi.org/10.1006/aphy.1994.1116}{{\em Annals Phys.} {\bfseries
  236} (1994) 374--391},
\href{http://arxiv.org/abs/hep-th/9308141}{{\ttfamily arXiv:hep-th/9308141
  [hep-th]}}.

\bibitem{York}
J.~W. York, ``Role of Conformal Three-Geometry in the Dynamics of
  Gravitation,'' \href{http://dx.doi.org/10.1103/PhysRevLett.28.1082}{{\em
  Phys. Rev. Lett.} {\bfseries 28} (Apr, 1972) 1082--1085}.

\bibitem{Hawking:1995fd}
S.~W. Hawking and G.~T. Horowitz, ``{The Gravitational Hamiltonian, action,
  entropy and surface terms},''
  \href{http://dx.doi.org/10.1088/0264-9381/13/6/017}{{\em Class. Quant. Grav.}
  {\bfseries 13} (1996) 1487--1498},
\href{http://arxiv.org/abs/gr-qc/9501014}{{\ttfamily arXiv:gr-qc/9501014
  [gr-qc]}}.

\bibitem{MaMa}
R.~B. Mann and D.~Marolf, ``{Holographic renormalization of asymptotically flat
  spacetimes},'' \href{http://dx.doi.org/10.1088/0264-9381/23/9/010}{{\em
  Class. Quant. Grav.} {\bfseries 23} (2006) 2927--2950},
\href{http://arxiv.org/abs/hep-th/0511096}{{\ttfamily arXiv:hep-th/0511096
  [hep-th]}}.

\bibitem{Hollands:2012sf}
S.~Hollands and R.~M. Wald, ``{Stability of Black Holes and Black Branes},''
  \href{http://dx.doi.org/10.1007/s00220-012-1638-1}{{\em Commun. Math. Phys.}
  {\bfseries 321} (2013) 629--680},
\href{http://arxiv.org/abs/1201.0463}{{\ttfamily arXiv:1201.0463 [gr-qc]}}.

\bibitem{Compere:2011ve}
G.~Compere and F.~Dehouck, ``{Relaxing the Parity Conditions of Asymptotically
  Flat Gravity},'' \href{http://dx.doi.org/10.1088/0264-9381/28/24/245016}{{\em
  Class. Quant. Grav.} {\bfseries 28} (2011) 245016},
  \href{http://arxiv.org/abs/1106.4045}{{\ttfamily arXiv:1106.4045 [hep-th]}}.
[Erratum: Class. Quant. Grav.30,039501(2013)].

\bibitem{Wald:1993nt}
R.~M. Wald, ``{Black hole entropy is the Noether charge},''
  \href{http://dx.doi.org/10.1103/PhysRevD.48.R3427}{{\em Phys. Rev.}
  {\bfseries D48} (1993) R3427--R3431},
\href{http://arxiv.org/abs/gr-qc/9307038}{{\ttfamily arXiv:gr-qc/9307038
  [gr-qc]}}.

\end{thebibliography}\endgroup

\end{document}